\documentclass[aip, jmp, onecolumn, groupedaddress,longbibliography, 12pt]{revtex4-1}
\usepackage{amsmath,amssymb}
\usepackage{graphicx}
\usepackage{color}
\usepackage{hyperref}
\begin{document}
\def\d{{\mathrm{d}}}
\def\lint{\hbox{\Large $\displaystyle\int$}}   
\def\hint{\hbox{\huge $\displaystyle\int$}}  
\title{\bf\Large Greybody factors for Myers--Perry black holes}
\author{\ \ \\ Petarpa Boonserm\,}
\email[]{petarpa.boonserm@gmail.com}
\affiliation{Department of Mathematics and Computer Science, Faculty of Science, \\
Chulalongkorn University, Phayathai Road, Patumwan, Bangkok 10330, Thailand\vskip1pt}
\author{Auttakit Chatrabhuti\,}
\email[]{dma3ac2@gmail.com}
\author{Tritos Ngampitipan\,}
\email[]{tritos.ngampitipan@gmail.com}
\affiliation{Department of Physics, Particle Physics Research Laboratory, Faculty of Science,\\
Chulalongkorn University, Phayathai Road, Patumwan, Bangkok 10330, Thailand\vskip1pt}
\author{Matt Visser\,}
\email[]{matt.visser@msor.vuw.ac.nz}
\affiliation{ \mbox{School of Mathematics, Statistics, and Operations Research,}\\
Victoria University of Wellington; \\
PO Box 600, Wellington 6140, New Zealand.\\}
\date{22 May 2014; \LaTeX-ed \today}

\begin{abstract}
Abstract:

\smallskip
\noindent
The Myers--Perry black holes are higher-dimensional generalizations of the usual (3+1)-dimensional rotating Kerr black hole.  They are of considerable interest in Kaluza--Klein models, specifically within the context of brane-world versions thereof. 
In the present article we shall consider the greybody factors associated with scalar field excitations of the Myers--Perry spacetimes, and develop some rigorous bounds on these greybody factors. These bounds are of relevance for characterizing both the higher-dimensional Hawking radiation, and the super-radiance, that is expected for these spacetimes. 

\bigskip
\noindent Keywords: 

\smallskip
\noindent
Hawking radiation, greybody factors, rigorous bounds, Myers--Perry black holes.

\end{abstract}
\maketitle
\clearpage
\bigskip
\hrule
\tableofcontents
\bigskip
\hrule
\clearpage
\section{Introduction.}

Greybody factors modulate the absorption cross-sections of classical black holes, and alter the closely related Hawking emission~\cite{Hawking:1974, Hawking:1975} probabilities of semi-classical black holes.~\cite{Page:1976a, Page:1976b, Bekenstein:1977, Escobedo:2008} Physically, the incoming or outgoing wave back-scatters off the  gravitational field surrounding the black hole, leading to a non-trivial transmission coefficient. In the case of Hawking radiation, this modifies the naive Planckian spectrum by multiplying it with a frequency-dependent greybody factor.  Explicitly evaluating these greybody factors is typically an impossible task, even for the simple case of the Schwarzschild black hole.~\cite{Schwarzschild} In view of this difficulty, techniques for placing analytic bounds on the greybody factors have now become of some interest.~\cite{Schwarzschild, Tritos:2012, Tritos:2013, Static-Spherically-Symmetric, Kerr-Newman} (Alternatively one might seek to extract qualitative or numerical information.~\cite{Ida, Creek1, Creek2})

The bounds developed in references \onlinecite{Schwarzschild}--\onlinecite{Kerr-Newman} apply to various black holes, 
(Schwarzschild, Reissner--Nordstr\"om, Kerr, Kerr--Newman, etcetera), 
and are all based on a very general technique for bounding one-dimensional barrier penetration probabilities; 
a technique that was first developed in reference \onlinecite{Visser}, with later formal developments to be found in references \onlinecite{miller-good}--\onlinecite{shabat-zakharov}, and additional related discussion in references \onlinecite{Boonserm:2010}--\onlinecite{Boonserm:2013}.  In the current article we shall apply the same sort of formalism to the Myers--Perry rotating black holes in (3+1+$n$) dimensions.~\cite{Myers-Perry, Emparan} 
The Myers--Perry black holes are particularly important in that they are the simplest of the higher-dimensional rotating black holes, being of particular interest in both Kaluza--Klein scenarios and in brane-world scenarios. 

We first describe the Myers-Perry spacetime,\cite{Myers-Perry, Emparan} setting up the relevant Teukolsky equation for scalar field excitations.~\cite{Press:1973} An important part of the technical analysis is the fact that we can place positivity constraints on both the separation constant and on the effective potential; without such positivity constraints progress would be severely limited. We then analyze both the greybody factors and (when relevant) super-radiant emission as a function of the angular momentum quantum number $m$. While zero angular momentum ($m=0$) serves as a good template for the other cases, there are some significant differences to take into account.  After completing the analysis and summarizing the general case, we specialize to (3+1) dimensions to verify compatibility with the usual Kerr black hole, and also consider the specific  (3+1+1)  five-dimensional case which is perhaps most relevant to brane-world models. We conclude with a brief discussion of the significance of our results.

\section{Teukolsky equation for scalar fields.}
 
In setting up the formalism, it is best to first focus on the geometry of the specific spacetimes under consideration, and then analyse the technical steps involved in separation of variables, leading up to the development of the Teukolsky equation for scalar field excitations. With this in hand, one can then proceed to examination of the effective potential. For some general background on black hole perturbation theory see references \onlinecite{Shibata:1995}--\onlinecite{Ferrari:2011}.

\subsection{Myers--Perry spacetime.}

The Myers--Perry geometry (with only one of the angular momentum parameters being non-zero) is described by the metric~\cite{Myers-Perry, Emparan}
\begin{eqnarray}
\d s^{2} &=& -\d t^{2} + \frac{\Sigma}{\Delta}\,\d r^2 + \Sigma \,\d\theta^2 + (r^2+a^2)\sin^2\theta \; \d\varphi^2
\nonumber\\
&& 
+{\mu\over r^{n-1}\Sigma}(\d t - a \sin^2\theta\;\d\varphi)^2
+ r^2\cos^2\theta \; \d\Omega^2_n.
\label{MPmetric}
\end{eqnarray}
Here
\begin{equation}
\Delta = r^2 + a^2 - \frac{\mu}{r^{n-1}}, \qquad\qquad  \Sigma = r^2 + a^2\cos^2\theta,
\label{Delta}
\end{equation}
and $\d\Omega_n^2$ is the line-element on the unit $n$-sphere $S^n$. We choose coordinates so that 
\begin{equation}
\d\Omega_n^2 = \d\theta_1^2 + \sin^2\theta_1 \;\d\theta_2^2 + \sin^2\theta_1\sin^2\theta_2 \; \d\theta_3^2 + \dots +
\left(\prod\nolimits_{i=1}^{n-1}\sin^2\theta_i\right) \d\theta_n^2,
\label{E:Sn-metric}
\end{equation}
whence recursively
\begin{equation}
\d\Omega_n^2(\theta_1,\dots,\theta_n) = \d\theta_1^2 + \sin^2\theta_1 \; \d\Omega_{n-1}^2(\theta_2,\dots,\theta_n).
\label{E:Sn-metric-recursive}
\end{equation}
(Several other coordinate conventions on the $n$-sphere are also relatively common.)
This Myers--Perry spacetime has $4+n$ dimensions, 4 of them ``usual'' and $n$ ``extra''. This is sometimes phrased as $3+1+n$ dimensions, (meaning 3 of space, 1 of time, and $n$ ``extra''  Kaluza--Klein dimensions). The black hole mass $M_{BH}$, and angular momentum $J$, are defined as follows
\begin{equation}
M_{BH} = \frac{(n+2)\,A_{n+2}}{16\pi G}\mu,\qquad\qquad J = \frac{2a}{n+2} \;M_{BH}.
\end{equation}
Here $G$ denotes the gravitational constant in the $(4+n)$-dimensional space-time, and the quantity $A_{n+2} = 2\pi^{(n+3)/2}/\Gamma[(n+3)/2] $ is the area of a $(n+2)$-dimensional unit sphere.   
The location of the black hole horizon $r_H$ is the solution of $\Delta(r_H) = 0$,  such that $\mu = r_H^{n-1}(r_H^2+a^2)$ is satisfied.
\begin{itemize}
\itemsep-2pt
\item In the specific case of $n=0$ this spacetime reduces to the standard Kerr black hole, 
with the usual inner and outer horizons.
\item In the specific case of $n=1$, we have $\mu = r_H^2+a^2$, so then  $r_H= \sqrt{\mu-a^2}$, and the horizon exists only when $a < \sqrt{\mu}$; in fact the horizon shrinks to zero area in the extreme limit $a \rightarrow \sqrt{\mu}$.  So the case $n=1$ is somewhat different from $n>1$. 
\item On the other hand, in the case of  $n\geqslant 2$, for $\mu>0$ a unique positive solution for $r_H$ always exists for all $a$. Indeed $r_H\in\left(0,\mu^{1/(n+1)}\right]$. 
\end{itemize}

\subsection{Separation of variables.}
In this article we will focus on scalar field emission from the Myers--Perry black hole.  The relevant excitations can be described by the Klein--Gordon equation
\begin{equation}
\partial_{\mu}\left(\sqrt{-g}\,g^{\mu\nu}\,\partial_{\nu} \Phi\right) = 0.
\label{scalar_eom}
\end{equation}
Here the metric determinant factorizes nicely into 4-dimensional and $n$-dimensional pieces. Specifically, with conventions as in equation (\ref{E:Sn-metric}), we have
\begin{equation}
\sqrt{-g} = \left(\Sigma\,\sin\theta\right) \times \left( r^n \cos^n\theta \right) \times \left(\prod_{i=1}^{n-1}\sin^{n-i}\theta_i\right),
\end{equation}
with the trailing factor arising from the unit $n$-sphere.

Similarly to the Kerr--Newman black hole in four dimensions, the Myers--Perry solution enjoys a hidden symmetry due to the existence of a Killing--Yano tensor.~\cite{Frolov}  In view of this, we can use the separation of variables ansatz~\cite{Carter:1968}
\begin{equation}
\Phi(t,r,\theta,\varphi,\theta_1,\dots,\theta_{n}) = 
e^{-i \omega t}\;e^{im\varphi}\;\tilde{R}_{j\ell m}(r)\;S_{\ell m}(\theta)\;Y_{jn}(\theta_1,\dots,\theta_{n}).
\end{equation}
Here the $Y_{jn}(\theta_1,\dots,\theta_n)$ are the quite standard hyper-spherical harmonics defined on the unit $n$-sphere, 
which satisfy the differential equation~\cite{Muller}
\begin{equation}
\Delta_{S^n} Y_{jn}(\theta_1,\dots,\theta_n)  + j(j+n-1)Y_{jn} = 0.
\end{equation}
The important observation is that for the $n$-sphere the Laplacian eigenvalues are $-j(j+n-1)$. In 4 dimensions ($n\to0$) these hyperspherical harmonics reduce to trivial constants, (and $j\to 0$). In 5 dimensions ($n\to1$) they are simply sines and cosines.
 If one wishes an explicit rendition of the Laplacian on the $n$-sphere then, with coordinates as in equation (\ref{E:Sn-metric}), we have
\begin{equation}
\sum_{k=1}^{n}\frac{1}{\prod_{i=1}^{n-1}\sin^{n-i}\theta_i}\partial_{\theta_k}
\left[\left(\prod_{i=1}^{n-1}\sin^{n-i}\theta_i \right)\frac{\partial_{\theta_k}Y_{jn}}{\prod_{i=1}^{k-1}\sin^2\theta_i} \right] 
+ j(j+n-1)Y_{jn} = 0.
\end{equation}
We mention in passing that when you choose coordinates to write the $n$-sphere metric recursively, as in equation (\ref{E:Sn-metric-recursive}),
then the Laplacian can also be expressed recursively
\begin{equation}
\Delta_{S^n} X 
=
{1\over \sin^{n-1}\theta_1} \; {\partial\over\partial{\theta_1}} \left( \sin^{n-1}\theta_1\;\; {\partial X\over\partial\theta_1}\right) 
+
{1\over\sin^2\theta_1} \; \Delta_{S^{n-1}} X.
\end{equation}

\bigskip
\noindent
In contrast to the hyper-spherical harmonics defined on the hyper-sphere $S^n$, the spheroidal harmonics $S_{\ell m}(\theta)\;e^{im\varphi}$ are defined on the two angular variables associated with the ``usual'' 4-dimensional part of the spacetime. They are the appropriate generalization of the standard spherical harmonics $Y_{\ell m}(\theta, \phi)$. The spheroidal harmonics satisfy the differential equation\cite{Ida}
\begin{eqnarray}
\left\{\frac{1}{\sin\theta\cos^{n}\theta}\frac{\d }{\d \theta}\left[\sin\theta\cos^{n}\theta\frac{\d }{\d \theta}\right] 
-  \left( \omega a\sin\theta - \frac{m}{\sin\theta} \right)^2  
- \frac{j(j + n - 1)}{\cos^{2}\theta}   
+ \lambda_{j\ell m}\right\}S_{\ell m}(\theta) = 0.
\nonumber\\
\label{ang2}
\end{eqnarray}
Note that going to 4 dimensions corresponds to setting $n\to 0$ \emph{and} setting $j\to 0$, in which case this differential equation reduces to that for the Kerr (or Kerr--Newman) geometry as given in reference~\onlinecite{Kerr-Newman}. These spheroidal harmonics are very closely related both to the Heun functions,~\cite{Heun:1, Heun:2, Heun:3, Heun:4} and to the hyper-spherical harmonics.~\cite{Muller, Fackerell:1977}

The separation constant $\lambda_{j\ell m}$ in this spheroidal differential equation is positive. To see this let us define a new variable by $\d u = \sin\theta\,\cos^{n}\theta\,\d \theta$, then
\begin{equation}
\frac{\d }{\d \theta} = \frac{\d u}{\d \theta}\frac{\d }{\d u} = \sin\theta\cos^{n}\theta\frac{\d }{\d u}.
\end{equation}
Therefore
\begin{equation}
\frac{1}{\sin\theta\cos^{n}\theta}\frac{\d }{\d \theta}\left[\sin\theta\cos^{n}\theta\frac{\d S(\theta)}{\d \theta}\right] = \frac{\d }{\d u}\left[\left(\sin\theta\cos^{n}\theta\right)^{2}\frac{\d S(\theta)}{\d u}\right].
\end{equation}
Then the angular equation (\ref{ang2}) for the spheroidal harmonics becomes
\begin{eqnarray}
\frac{\d }{\d u}\left[\left(\sin\theta\cos^{n}\theta\right)^{2}\frac{\d S(\theta)}{\d u}\right] &=& \left[ \left( \omega a\sin\theta - \frac{m}{\sin\theta} \right)^2  
+ \frac{j(j + n - 1)}{\cos^{2}\theta}   
- \lambda_{j\ell m}\right]S(\theta).
\end{eqnarray}
Multiplying the above equation by $S(\theta)$ 
and integrating both sides over $u$ yields
\begin{eqnarray}
&&\int S(\theta)\frac{\d }{\d u}\left[\left(\sin\theta\cos^{n}\theta\right)^{2}\frac{\d S(\theta)}{\d u}\right]\d u 
\nonumber\\
&& \qquad =
\int\left[
\left( \omega a\sin\theta - \frac{m}{\sin\theta} \right)^2  
+ \frac{j(j + n - 1)}{\cos^{2}\theta}   
- \lambda_{j\ell m}
\right]S^{2}(\theta)\;\d u.
\label{integrated}
\end{eqnarray}
Integrate the left hand side by parts, using periodicity to discard boundary terms, and then rearrange to obtain
\begin{eqnarray}
\lambda_{j\ell m} \int S^2(\theta) \; \d u &=& \int\left[
\left( \omega a\sin\theta - \frac{m}{\sin\theta} \right)^2  + \frac{j(j + n - 1)}{\cos^{2}\theta} \right]S^{2}(\theta)\; \d u 
\nonumber\\
&& + \int  \left[\left(\sin\theta\cos^{n}\theta\right)^{2} \left(\frac{\d S(\theta)}{\d u}\right)^2 \right]   \d u.
\end{eqnarray}
Now the right hand side of this equation is manifestly positive, as is the factor $\int S^2 \, \d u $ on the left hand side. Therefore the separation constant $\lambda_{j\ell m}$ is guaranteed to be positive. 

\subsection{Effective potential.}

We now construct the effective potential, starting from the radial part of the variable-separated Klein--Gordon equation.\cite{Ida,Creek1,Creek2}
We have
\begin{equation}
\left\{\frac{1}{r^n}\frac{\d}{\d r}\left[r^n\Delta\frac{\d}{\d r}\right]+\frac{[(r^2 + a^2)\omega - m a]^2}{\Delta} -\frac{j(j+n-1)a^2}{r^2} -\lambda_{j\ell m}\right\}\tilde{R}_{j\ell m}(r)=0.
\end{equation}
Let us now define a new radial mode function
\begin{equation}
\tilde{R}_{j\ell m}(r) =\frac{r^{-\frac{n}{2}}R_{j\ell m}(r)}{\sqrt{r^2+a^2}}.
\end{equation}
It is now a quite standard calculation to show that the radial Teukolsky equation, (the Regge--Wheeler-like equation governing the radial modes), is given by~\cite{Ida,Creek1,Creek2}
\begin{equation}
\left\{\frac{\d ^{2}}{\d r_{*}^{2}} - U_{j\ell m}(r)\right\}R_{j\ell m}(r) = 0,
\end{equation}
where $r_{*}$ is the standard ``tortoise coordinate"
\begin{equation}
\d r_{*} = \frac{r^{2} + a^{2}}{\Delta(r)}\; \d r.
\end{equation}
Note that the tortoise coordinates can be expressed as
\begin{equation}
r_* = \int_{r_H}^r \frac{r^2+a^2}{\Delta(r)}dr \sim A_n \ln(r-r_H) + B_{n}(r),
\end{equation}
where the exact expressions for the coefficients $A_n$ and functions $B_n(r)$ depend on the number of extra dimensions $n$.  However, we can quite generally observe that as $r \rightarrow r_H$ we have $r_* \rightarrow - \infty$, and as $r \rightarrow \infty$ we have $r_* \rightarrow \infty$.  So the region $r>r_H$ outside the black hole, (the domain of outer communication), maps into the entire real line $-\infty\leq r_* \leq +\infty$ in terms of the tortoise coordinate. 

\noindent
The Teukolsky potential, (sometimes called the Regge--Wheeler--Teukolsky potential), is now seen to be
\begin{eqnarray}
U_{j\ell m}(r) &=& \frac{\Delta(r)}{\left(r^{2} + a^{2}\right)^{2}}\left[\lambda_{j\ell m} + \frac{j(j+n-1)a^2}{r^2} + \frac{n(n-2)\Delta(r)}{4r^2}
+\frac{n\Delta^{\prime}(r)}{2r}  \right.  \nonumber\\
&&\left. \qquad\qquad\qquad
 -\frac{3r^2\Delta(r)}{(r^2+a^2)^2} +\frac{[r\Delta(r)]^{\prime}}{r^2+a^2}\right]  
 - \left( \omega - \frac{m a}{r^{2} + a^{2}} \right)^2. \label{V(r)}
\end{eqnarray}
Note that for  $j=n=0$ this reduces to the Teukolsky potential for the ordinary  Kerr black hole in 4 dimensional space-time. (See reference \onlinecite{Kerr-Newman}.)
For purposes of calculation, we now define quantities
\begin{equation}
\varpi(r) = \frac{a}{a^2+r^2},
\end{equation}
and more specifically
\begin{equation}
\Omega_H = \frac{a}{a^2+r_H^2}.
\end{equation}
Here $\varpi(r)$ is related to frame dragging, while $\Omega_H$ is the ``angular velocity'' of the event horizon.\cite{Kerr-Newman}  We can now re-express the Teukolsky potential as
\begin{equation}
U_{j\ell m}(r) = V_{j\ell m}(r) - (\omega - m\varpi)^2,
\end{equation}
with
\begin{eqnarray}
V_{j\ell m}(r) &=& \frac{\Delta(r)}{\left(r^{2} + a^{2}\right)^{2}}\left[\lambda_{j\ell m} + \frac{j(j+n-1)a^2}{r^2} 
\right.  \nonumber\\
&&
\qquad\qquad\qquad
\left. 
+ \frac{n(n-2)\Delta(r)}{4r^2}+\frac{n\Delta^{\prime}(r)}{2r} 
-\frac{3r^2\Delta(r)}{(r^2+a^2)^2} +\frac{[r\Delta(r)]^{\prime}}{r^2+a^2}\right].
\end{eqnarray}

\subsection{Positivity properties.}
To show positivity of $V_{j\ell m}(r)$, we start by noting that $\Delta(r) >0$ outside the horizon, (that is for $r>r_H$).  This is standard for $n=0$, and trivial for $n=1$. For $n\geq 1$ we generically re-express $\Delta(r)$ as
\begin{eqnarray}
\Delta(r) &=& r^2+a^2 -r^{1-n}\mu \nonumber\\
               &=& r^2+a^2 -\left({r}/{r_H}\right)^{1-n} (r_H^2 + a^2)\nonumber\\
               &\geq& (r_H^2 +a^2)\left(1-\left(r_H/r\right)^{n-1}\right).
\end{eqnarray}
Since $r \geq r_H$, we can see that $\Delta(r) \geq 0$ for $n \geq 1$.  
Using this result, we make the following observations.  First,  for $n\geq 1$ we have
\begin{eqnarray}
\frac{[r\Delta(r)]^{\prime}}{r^2+a^2}-\frac{3r^2\Delta(r)}{(r^2+a^2)^2}  & \propto & [r\Delta(r)]^{\prime}(r^2+a^2) - 3r^2\Delta(r) \nonumber\\
& = & a^2(r^2+a^2) + \frac{\mu}{r^{n-1}} \left[(n+1)r^2 + (n-2)a^2 \right] \nonumber\\
&=& a^2\Delta(r) + \frac{\mu}{r^{n-1}} \left[(n+1)r^2 + (n-1)a^2 \right]\nonumber\\
&\geq& 0.
\end{eqnarray}
Note that the equivalent result for $n=0$ was already derived in reference \onlinecite{Kerr-Newman} for the Kerr--Newman spacetime. Second,  for $n \geq 0$, we also have
\begin{eqnarray}
 \frac{n(n-2)\Delta(r)}{4r^2}+\frac{n\Delta^{\prime}(r)}{2r}  
 &\propto& n \{ (n-2)\Delta(r)  + 2  r\Delta^{\prime}(r)  \}
 \nonumber\\
 &=& n\{ (n+2)r^2 + (n-2)a^2 + n\mu r^{1-n} \}.
\end{eqnarray}
Now for $n\geq 2$ this quantity is certainly positive. For $n=0$ this quantity is identically zero. For $n=1$ this quantity reduces to $3r^2-a^2+\mu=3 r^2-r_H^2 \geq 0$ (provided the horizon exists).  In all situations the relevant quantity is non-negative.
 Thus, by now combining these results with the fact that $\lambda_{j\ell m} > 0$, and the fact that both $n \geq 0$ and $j\geq0$, we can conclude that $V_{j\ell m}(r)$ is always positive for all values of $j$, $\ell$, $m$, and $r$.

\subsection{Super-radiance.}

Now note that the effective potential is
\begin{equation}
U_{j\ell m}(r) = V_{j\ell m}(r) - (\omega - m\varpi)^2; \qquad \qquad V_{j\ell m}(r) \geq 0.
\end{equation}
However, the quantity $\omega - m\varpi$ can under suitable circumstances  change sign. This is the harbinger of super-radiance. 
Some rather general analyses can be found in references \onlinecite{Richartz:2009}--\onlinecite{Zeldovich}, while a specific analysis  closely related  to the current situation can be found in reference \onlinecite{Kerr-Newman}. 
The key point is that super-radiance is a phenomenon in which the reflected wave is larger in its amplitude than the incident wave. From mathematical point of view, super-radiance is a phenomenon in which $|r| > 1$, where $r$ is the reflection coefficient. 
Super-radiance will occur once $\omega - m\varpi$  changes sign in the domain of outer communication which, given the asymptotic behaviour of $\varpi$, occurs whenever $0 < \omega < m \Omega_H$, that is $m > m_* \equiv \omega/\Omega_H$. 
Once super-radiance occurs, the bound on the greybody factor becomes a bound on the spontaneous emission amplitude. A detailed discussion of this particular issue can be found in reference~\onlinecite{Kerr-Newman}.

\section{Analytic bound for scalar transmission.}
From reference \onlinecite{Visser}, (see also references \onlinecite{miller-good}, \onlinecite{bogoliubov}, \onlinecite{analytic}, and \onlinecite{shabat-zakharov} for further developments and applications), we have the extremely general result that
\begin{equation}
T_{j\ell m} \geq \text{sech}^{2}\left(\int_{-\infty}^{\infty}\vartheta \, \d r_{*}\right),\label{T}
\end{equation}
where
\begin{equation}
\vartheta = \frac{\sqrt{[h'(r_*)]^{2} + \left[U_{j\ell m}(r_*) + h^{2}(r_*)\right]^{2}}}{2h(r_*)},\label{theta}
\end{equation}
for any positive function ${h}(r_*)$.  Equivalently
\begin{equation}
\vartheta = \frac{\sqrt{[h'(r_*)]^{2} + \left[V_{j\ell m}(r_*) - (\omega-m\varpi)^2 + h^{2}(r_*)\right]^{2}}}{2h(r_*)}.\label{theta2}
\end{equation}
We shall now use the positivity properties of $\lambda_{j\ell m}$ and $V_{j\ell m}$, together with the super-radiant/ non-super-radiant distinction, to systematically analyse this bound in various cases. In particular
\begin{itemize}
\item The modes $m< {m_*} \equiv \omega/\Omega_H$ are not super-radiant.
\item The modes $m \geq {m_*} \equiv \omega/\Omega_H$ are super-radiant.
\end{itemize}
In situations where super-radiance occurs, in addition to the greybody factor $T_{j\ell m}$, there is a closely related spontaneous emission rate which satisfies the bound~\cite{Kerr-Newman}
\begin{equation}
\Gamma_{j\ell m} \leq \omega \sinh^2\left(\int_{-\infty}^{\infty}\vartheta \, \d r_{*}\right).\label{R}
\end{equation}

\section{Non-super-radiant modes ($m < m_*$).}

It is convenient to split the discussion of non-super-radiant modes into three sub-cases:
\begin{itemize}
\item $m=0$ zero-angular-momentum modes: This is the most fundamental case, and most straightforward case to analyze. This case provides a useful template for the more complicated situations. 
\item$m\neq0$ nonzero-angular-momentum modes:  These are most conveniently further split into two sub-cases.
    \begin{itemize}
    \item $m < 0$ negative-angular-momentum modes.
    \item $m\in(0,{m_*})$ low-lying positive-angular-momentum modes.
    \end{itemize}
\end{itemize}

\subsection{Zero angular momentum modes ($m=0$).}

We choose $\tilde{h}(r_*)=\omega >0$ and $m=0$, then
\begin{eqnarray}
U_{j\ell,m=0}(r) &=& \frac{\Delta(r)}{\left(r^{2} + a^{2}\right)^{2}}\left[\lambda_{j\ell,m=0} + \frac{j(j+n-1)a^2}{r^2} + \frac{n(n-2)\Delta(r)}{4r^2}   \right.  \nonumber\\
&& \left. +\frac{n\Delta^{\prime}(r)}{2r} -\frac{3r^2\Delta(r)}{(r^2+a^2)^2} +\frac{[r\Delta(r)]^{\prime}}{r^2+a^2}\right]  - \omega^2. \label{Vm=0}
\end{eqnarray}
Then
\begin{eqnarray}
T &\geq& \text{sech}^{2}\left(\frac{1}{2\omega}\int_{-\infty}^{\infty}\left |V \right |\d r_{*}\right)\nonumber\\
  &=&    \text{sech}^{2}\left(\frac{1}{2\omega}\int_{r_{h}}^{\infty}\left | V(r) \right | \frac{r^{2} + a^{2}}{\Delta(r)}\d r\right)\nonumber\\
  &=&    \text{sech}^{2}\left[\frac{1}{2\omega}\int_{r_{h}}^{\infty}\left | \frac{1}{r^{2} + a^{2}}\left\{\lambda_{j\ell,m=0} + \frac{j(j+n-1)a^2}{r^2} + \frac{n(n-2)\Delta(r)}{4r^2} \right.\right. \right.\nonumber\\
  && \qquad\qquad
  \left.\left.\left. -\frac{3r^2\Delta(r)}{(r^2+a^2)^2} +\frac{n\Delta^{\prime}(r)}{2r}+\frac{[r\Delta(r)]^{\prime}}{r^2+a^2}\right\} \right| \d r\right].
\end{eqnarray}
For $n \geq 1$ and $r \geq r_H$, in view of the positivity properties of the separation constant and effective potential, we can replace $\int |\cdots|\,\d r \rightarrow |\int \cdots \d r|$.  Therefore
\begin{eqnarray}
T &\geq& \text{sech}^{2}\left|\frac{1}{2\omega}\int_{r_{h}}^{\infty}\frac{1}{r^{2} + a^{2}}\left\{\lambda_{j\ell,m=0} + \frac{j(j+n-1)a^2}{r^2} + \frac{n(n-2)\Delta(r)}{4r^2} 
\right.\right. 
\nonumber\\
&&\qquad\qquad \left.\left. 
  -\frac{3r^2\Delta(r)}{(r^2+a^2)^2} +\frac{n\Delta^{\prime}(r)}{2r}+\frac{[r\Delta(r)]^{\prime}}{r^2+a^2}\right\}  \d r\right|. 
  \label{boundn}
\end{eqnarray}
We would like to integrate this equation term by term.  Start by considering the first term:
\begin{eqnarray}
 \int_{r_H}^{\infty} \frac{\lambda_{j\ell,m=0}}{r^2+a^2}\, \d r &=& 
 \left.\frac{\lambda_{j\ell,m=0}}{a}\arctan \frac{r}{a} \right|_{r_H}^{\infty} 
= 
 \frac{\lambda_{j\ell,m=0}}{a} \arctan \frac{a}{r_H}.
 \label{Lambda_integral}
\end{eqnarray}
For the last two integrals, we can show that they can be simplified as follows:
\begin{equation}
\int_{r_H}^{\infty} \frac{1}{r^2+a^2}\left[  -\frac{3r^2\Delta(r)}{(r^2+a^2)^2} +\frac{[r\Delta(r)]^{\prime}}{r^2+a^2}\right]\,\d r = \int_{r_H}^{\infty} \frac{r^2\Delta(r)}{(r^2+a^2)^3}\;\d r.
\end{equation}
This can be explicitly integrated (for instance by using {\sf Mathematica}) and we arrive at
\begin{eqnarray}
\int_{r_H}^{\infty} \frac{r^2\Delta(r)}{(r^2+a^2)^3}\d r &=& 
\frac{n}{8r_H}- \frac{n(n-2)(r_H^2+a^2)}{8(n+2)r_H^3} \; \;{}_2F_1\left(1,\frac{n+2}{2},\frac{n+4}{2},-\frac{a^2}{r_H^2}\right)
\nonumber\\
&-&
 \frac{a^2}{4r_H(r_H^2+a^2)}+\frac{1}{2a}\arctan \frac{a}{r_H}.
\label{last_2_terms}
\end{eqnarray}
Here ${}_2F_1(z_1,z_2,z_3,z_4)$ is the hypergeometric function.   Let us now consider the $j$-dependent integral:
\begin{equation}
 \int_{r_H}^{\infty} \frac{j(j+n-1)a^2}{r^2(r^2+a^2)} dr = \frac{j(j+n-1)}{r_H}-\frac{j(j+n-1)}{a}\arctan \frac{a}{r_H}.
 \label{j-integral}
 \end{equation}
We can also integrate the $n$-dependent terms as
\begin{eqnarray}
 \int_{r_H}^{\infty}\frac{1}{r^2+a^2}\left[\frac{n(n-2)\Delta(r)}{4r^2}  +\frac{n\Delta^{\prime}(r)}{2r}\right] dr &=&  
 \frac{n^2(r_H^2+a^2)}{4(n+2)r_H^3} \; {}_2F_1\left(1,\frac{n+2}{2},\frac{n+4}{2},-\frac{a^2}{r_H^2}\right)
 \nonumber\\
 &+&  \frac{n(n-2)}{4r_H}+\frac{n}{a}\arctan \frac{a}{r_H}.
\label{n-integral}
\end{eqnarray}
Finally, combining the results from equation (\ref{Lambda_integral}), (\ref{last_2_terms}), (\ref{j-integral}), and (\ref{n-integral}), we obtain
\begin{equation}
T_{j\ell,m=0} \geq \text{sech}^{2}\left|\frac{1}{2\omega r_H} I_{j\ell,m=0} \right|,
\end{equation}
where we define
\begin{eqnarray}
I_{j\ell,m=0} &=& \frac{n(2n-3)}{8}+j(j+n-1)+\frac{a^2}{4(r_H^2+a^2)} \nonumber\\
&& + \left(\frac{2n+1}{2} - j(j+n-1) + \lambda_{j\ell,m=0}\right)\frac{r_H}{a}\arctan \frac{a}{r_H}\nonumber\\
&& + \frac{n(r_H^2+a^2)}{8r_H^2} \;\; {}_2F_1\left(1,\frac{n+2}{2},\frac{n+4}{2},-\frac{a^2}{r_H^2}\right). \label{def_I}
\end{eqnarray}
For a consistency check, consider the limit $a\rightarrow 0$ (with both $n=0$ and $j=0$),
\begin{eqnarray}
\lim_{a\rightarrow 0} I_{j=0,\ell,m=0} &=& \lim_{a\rightarrow 0}\left[ -\frac{a^2}{4(r_H^2+a^2)}+\left(  \frac{1}{2}+ \lambda_{j=0,\ell,m=0} \right) \frac{r_H}{a} \arctan \frac{a}{r_H}\right]\nonumber\\
&=& \frac{1}{2} + \lambda_{j=0,\ell,m=0}.
\end{eqnarray}
This is the same result as for the Kerr black hole, (the Kerr--Newman black hole for $Q=0$),
as is to be expected.

\subsection{Non-zero angular momentum mode ($m \neq 0$).}

From the basic inequality we have
\begin{equation}
T_{jlm} \geq \text{sech}^{2}\left[\hint_{\!\!\!\!\!\!\!\!\!\!-\infty}^{\infty} \frac{\sqrt{[\tilde{h}'(r_*)]^{2} + \left[\tilde{U}_{j\ell m}(r_*) + \tilde{h}^{2}(r_*)\right]^{2}}}{2\tilde{h}(r_*)}\d r_{*}\right],\label{T_m_neq_0}
\end{equation}
for all $\tilde{h}(r_*)>0$.  By now using the triangle inequality
\begin{equation}
|a|+|b| \geq \sqrt{a^2+b^2},
\end{equation}
we have
\begin{eqnarray}
T_{jlm} 
&\geq& 
\text{sech}^{2}\left[\lint_{\!\!\!\!\!\!-\infty}^{\infty} \frac{\left|\tilde{h}'(r_*)\right| + \left|\tilde{U}_{j\ell m}(r_*) + \tilde{h}^{2}(r_*)\right|}{2\tilde{h}(r_*)}\d r_{*}\right]
 \nonumber\\
&\geq& 
\text{sech}^{2}\left[\lint_{\!\!\!\!\!\!-\infty}^{\infty} \frac{\left|\tilde{h}'(r_*)\right| }{2\tilde{h}(r_*)}\d r_{*} 
+ \lint_{\!\!\!\!\!\!-\infty}^{\infty} \frac{\left|\tilde{U}_{j\ell m}(r_*) + \tilde{h}^{2}(r_*)\right|}{2\tilde{h}(r_*)}\d r_{*} \right].
\end{eqnarray}
Provided that $\tilde{h}^\prime(r_*)$ is monotone, we have
\begin{equation}
\int_{-\infty}^{\infty} \frac{|\tilde{h}^\prime(r_*)|}{2\tilde{h}(r_*)}\d r_*  = 
\left\{\begin{array}{cc} 
\frac{1}{2} \ln \frac{\tilde{h}(\infty)}{\tilde{h}(-\infty)} & \text{for } \tilde{h}^\prime(r_*) > 0; \\ 
&  \\ 
-\frac{1}{2}  \ln \frac{\tilde{h}(\infty)}{\tilde{h}(-\infty)} & \text{for } \tilde{h}^\prime(r_*) < 0.
\end{array}\right. 
\end{equation}
Let us now rewrite the potential as
\begin{equation}
U_{j\ell m} = V_{j\ell,m}  - \left(\omega - m \; \varpi(r) \right)^2.
\end{equation}
This form of potential is exactly the same as for the 4-dimensional Kerr--Newman black hole, and thus we simply choose
\begin{equation}
\tilde{h}(r_*) = h(r) = \omega - m \varpi(r).
\label{h_def1}
\end{equation}
Note that this choice for $h(r)$ is always monotonic as a function of $r$.  However, we can see that $h(r)$ is positive if and only if $\omega > m \Omega_H$.  This condition is satisfied for $m < \omega/\Omega_H$, (that is $m < m_*$), where the mode does not suffer from super-radiant instability.

\subsubsection{Negative-angular-momentum modes ($m<0$).}

Note that in this case, for $h(r)$ defined in equation (\ref{h_def1}),
\begin{equation}
\frac{\tilde{h}(\infty)}{\tilde{h}(-\infty)}= \frac{h(\infty)}{h(r_H)} = \frac{\omega}{\omega-m\Omega_H} = \frac{1}{1-m\Omega_H / \omega} < 1.
\end{equation}
Then
\begin{equation}
\frac{1}{2} \left| \ln \left[ \frac{\tilde{h}(\infty)}{\tilde{h}(-\infty)}\right] \right| = \frac{1}{2} \ln (1-m \Omega_H/\omega).
\end{equation}
Note also that in this case we have $\omega - m \Omega_H > h(r) > \omega$, so
\begin{equation}
\int_{-\infty}^{\infty} \frac{|U_{j \ell m}+h^2(r)|}{2h(r)} \d r_*  = \int_{-\infty}^{\infty} \frac{|V_{j\ell m}|}{2h(r)} \d r_* <\int_{-\infty}^{\infty} \frac{V_{j\ell m}}{2\omega} \d r_* .
\end{equation}
Then
\begin{eqnarray}
T_{j\ell,m<0} &\geq& \text{sech}^2 \left\{ \frac{1}{2} \ln(1-m\Omega_H/\omega) + \int_{-\infty}^{\infty} \frac{V_{j \ell,m<0}}{2\omega} \d r_* \right\} ,\\
&\geq& \text{sech}^2\left\{ \frac{1}{2} \ln(1-m/m_*) + \frac{1}{2\omega r_H} I_{j \ell,m<0}\right\}.
\end{eqnarray}
It is easy to see that this result is very similar to the result we have for $m=0$,  with the replacement $\lambda_{j\ell, m=0} \rightarrow \lambda_{j\ell, m<0} $.  We can write down $I_{j\ell m}$ explicitly as
\begin{eqnarray}
I_{j\ell m} &=& \frac{n(2n-3)}{8}+j(j+n-1)+\frac{a^2}{4(r_H^2+a^2)} 
\nonumber\\
&& + \left(\frac{2n+1}{2} - j(j+n-1) + \lambda_{j\ell m}(a\omega)\right)\frac{r_H}{a}\arctan \frac{a}{r_H}\nonumber\\
&& + \frac{n(r_H^2+a^2)}{8r_H^2} \; {}_2F_1\left(1,\frac{n+2}{2},\frac{n+4}{2},-\frac{a^2}{r_H^2}\right).
\label{def_Im}
\end{eqnarray}

\subsubsection{Low-lying positive-angular-momentum modes $(m \in (0,m_*))$.}

Recall that for $m_*> m  >0$, $h(r)$ is positive and monotonic as a function of $r$, for this situation we first consider
\begin{equation}
\frac{\tilde{h}(\infty)}{\tilde{h}(-\infty)} = \frac{h(\infty)}{h(r_H)} = \frac{\omega}{\omega-m\Omega_H} = \frac{1}{1-m\Omega_H/\omega}> 1.
\end{equation}
Then, we have
\begin{equation}
\frac{1}{2} \left| \ln \left[ \frac{\tilde{h}(\infty)}{\tilde{h}(-\infty)}\right] \right| = - \frac{1}{2} \ln (1-m \Omega_H/\omega).
\end{equation}
Note also that in this case we have $\omega - m \Omega_H < h(r) < \omega$, so
\begin{equation}
\int_{-\infty}^{\infty} \frac{|U_{j \ell m}+h^2(r)|}{2h(r)} \d r_*  = \int_{-\infty}^{\infty} \frac{|V_{j\ell, m>0}|}{2h(r)} \d r_* <\int_{-\infty}^{\infty} \frac{V_{j\ell, m>0}}{2(\omega-m\Omega_H)} \d r_* .
\end{equation}
Then, we arrive at the result
\begin{eqnarray}
T_{j\ell,m>0} &\geq& \text{sech}^2 \left\{ -  \frac{1}{2} \ln(1-m\Omega_H/\omega) + \int_{-\infty}^{\infty} \frac{V_{j \ell,m>0}}{2(\omega-m\Omega_H)} \d r_* \right\} , \\
&\geq& \text{sech}^2\left\{ - \frac{1}{2} \ln(1-m/m_*) + \frac{1}{2r_H\omega(1-m/m_*)} I_{j \ell,m>0}\right\} ,
\end{eqnarray}
where $I_{j\ell, m>0}$ is defined by equation (\ref{def_Im}).

\section{Super-radiant modes $(m \geq m_*)$.}

It is a good strategy to split the super-radiant modes into two sub-classes depending on the relative sizes of $\omega^2$ and $(\omega - m\Omega_H)^2$.  Note that $\omega^2 = (\omega - m\Omega_H)^2$ when $m = 2\omega/\Omega_H = 2m_*$.  This suggests that it might be useful to split the super-radiant modes as follows:
\begin{itemize}
  \item $m \in [m_*, 2m_*)$.
  \item $m \in [2m_*,\infty)$.
\end{itemize}

\subsection{Low-lying super-radiant modes $\left(m \in [m_*,2m_*)\right)$.}

In this region we have $\omega^2 > (\omega - m \Omega_H)^2$ and we choose
\begin{equation}
h(r) = \text{max}\left\{\omega - m \varpi(r), m\Omega_H - \omega\right\}.
\label{h_low}
\end{equation}
We can see that $h(r) > 0$ and monotone decreasing as we move from spatial infinity to the horizon, and become a flat horizontal line near the horizon.  Note that $h(r) \geq m\Omega_H - \omega$ everywhere.   By using $h(r)$ as defined in equation (\ref{h_low}), we have
\begin{equation}
\int_{-\infty}^{\infty} \frac{|h^\prime(r)|}{h(r)} \d r_*  = |\ln h(r)|_{r_H}^{\infty} = \ln\left( \frac{\omega}{m\Omega_H - \omega}\right) = - \ln(m/m_* - 1).
\end{equation}
It is now straightforward to show that
\begin{equation}
\int_{-\infty}^{\infty} \frac{V_{j\ell m}}{2h(r)} \d r_*  \leq \int_{-\infty}^{\infty} \frac{V_{j\ell m}}{2(m\Omega_H-\omega)} \d r_*  '
= \frac{I_{j\ell m}}{2(m\Omega_H-\omega)r_H} = \frac{I_{j\ell m}}{2\omega(m/m_*-1)r_H},
\end{equation}
where $I_{j\ell m}$ is defined in equation (\ref{def_Im}).  The last integral we need to perform is
\begin{equation}
J_m^{\text{low}} = \int_{-\infty}^{\infty} \frac{h(r)^2 - (\omega - m \varpi(r))^2}{2h(r)} \,\d r_* .
\end{equation}
Note that with our choice of $h(r)$, the integrand in above integral is zero over much of the relevant range.  To be more precise, we are interested only in
\begin{equation}
J_{m}^{\text{low}} = \int_{r_H}^{r_0} \frac{(\omega - m\Omega_H)^2 - (\omega - m \varpi(r))^2}{2(m\Omega_H-\omega)} \frac{r^2+a^2}{\Delta}\,\d r.
\end{equation}
The upper limit of integration $r_0$ is defined by the condition
\begin{equation}
m\left[\Omega_H + \varpi(r_0) \right] = 2\omega,
\end{equation}
or we can write down $r_0$ explicitly as
\begin{equation}
r_0 = \sqrt{r_H^2 + \frac{2(m-m_*)}{2m_*-m}(r_H^2+a^2)}.
\end{equation}
Notice that the upper limit $r_0 > r_H$ for $m \in [m_*,2m_*)$.  Then
\begin{equation}
J_{m}^{\text{low}} = \frac{m}{2(m\Omega_H-\omega)} 
\int_{r_H}^{r_0}\left(\Omega_H - \varpi(r)\right)
\left(m\varpi(r)+m\Omega_H-2\omega\right) \frac{r^2+a^2}{\Delta}\,\d r.
\end{equation}
However, for the relevant domain of integration we have
\begin{equation}
0\leq \left(m\varpi(r)+m\Omega_H-2\omega\right) \leq 2(m\Omega_H-\omega).
\end{equation}
Then we can conclude that
\begin{equation}
J_{m}^{\text{low}} \leq m \int_{r_H}^{r_0} \left(\Omega_H - \varpi(r)\right) \frac{r^2+a^2}{\Delta}\,\d r 
= m\Omega_H \int_{r_H}^{r_0}\frac{r^{n-1}(r-r_H)(r+r_H)}{r^{n-1}(r^2+a^2)-r_H^{n-1}(r_H^2+a^2)}\,\d r.
\label{j_low}
\end{equation}
This integral is finite, and one can evaluate it exactly for each value of $n$.  (The \emph{integrand} is in fact finite as $r\to r_H$ by the l'H\^opital rule.) By now combining all these results, we have
\begin{equation}
T_{j\ell,m\in[m_*,2m_*)} \geq \text{sech}^2\left\{-\frac{1}{2}\ln(m/m_*-1) + \frac{I_{j\ell,m\in[m_*,2m_*)}}{2r_H\omega(m/m_*-1)} + J^{\text{low}}_m \right\}.
\end{equation}

\subsection{Highly super-radiant modes $(m \geq 2m_*).$}

In this region we have $(\omega - m\Omega_H)^2 > \omega^2$, so we can choose
\begin{equation}
h(r) = \text{max}\left\{m \varpi(r) - \omega,\omega\right\}.
\label{h_high}
\end{equation}
It is not difficult to see that $h(r)$ is both positive and monotone decreasing as we move from the horizon to spatial infinity.  Note also that $h(r) \geq \omega$ for the relevant domain.  By using equation (\ref{h_high}), we have
\begin{equation}
\int_{-\infty}^{\infty} \frac{|h^\prime(r)|}{h(r)} \,\d r_*  = |\ln h(r)|_{r_H}^{\infty} = \ln\left( \frac{m\Omega_H - \omega}{\omega}\right) =  \ln(m/m_* - 1).
\end{equation}
We also obtain
\begin{equation}
\int_{-\infty}^{\infty} \frac{V_{j\ell m}}{2h(r)} \, \d r_*  \leq \int_{-\infty}^{\infty} \frac{V_{j\ell m}}{2\omega} \d r_*  = \frac{I_{j\ell m}}{2\omega r_H},
\end{equation}
where $I_{j\ell m}$ is defined in equation (\ref{def_Im}) as for the previous cases.  Finally, we are left with the integral
\begin{equation}
J_m^{\text{high}} = \int_{-\infty}^{\infty} \frac{h(r)^2 - (\omega - m \varpi(r))^2}{2h(r)} \, \d r_* .
\end{equation}
Again the integrand is zero over much of the domain of integration.  That is, we are only interested in
\begin{equation}
J_m^{\text{high}} = \int_{r_0}^{\infty} \frac{\omega^2 - (\omega - m \varpi(r))^2}{2\omega} \frac{r^2+a^2}{\Delta}\,\d r.
\label{j_high}
\end{equation}
Here the lower bound of integration, $r_0$, is now defined by
\begin{equation}
m\varpi(r_0) = 2\omega,
\end{equation}
implying
\begin{equation}
r_0 = a \sqrt{\frac{m}{2\omega a} -1}.
\end{equation}
Recall that $m \geq 2m_*$ in this region, we have
\begin{equation}
r_0 \geq a \sqrt{\frac{m_*}{\omega a} -1} = a \sqrt{\frac{r_H^2+a^2}{a^2} -1} = r_H.
\end{equation}
The integral $J_m^{\text{high}}$ is finite.  (In fact, the \emph{integrand} is finite as $r\to r_0$, and falls of as $1/r^2$ as $r\to\infty$.) After assembling all results we have, we finally obtain
\begin{equation}
T_{j\ell,m \geq 2m_*} \geq \text{sech}^2\left\{\frac{1}{2}\ln(m/m_*-1) + \frac{I_{j\ell, m\geq 2 m_*}}{2r_H\omega} + J^{\text{high}}_m \right\}.
\end{equation}

\section{Summary of the general case.}

Collecting the results for the low-lying and highly super-radiant modes, together with the non-super-radiant modes, we have the following bounds for the transmission probabilities:
\begin{equation}
T_{j \ell m} \geq \left\{\begin{array}{ll}
\text{sech}^2\left\{ \frac{1}{2} \ln(1-m/m_*) + \frac{1}{2 r_H \omega} I_{j \ell m}\right\} 
& \text{for } m < 0; 
\\ \\ 
\text{sech}^{2}\left\{\frac{1}{2r_H \omega} I_{j\ell m} \right\} & \text{for } m = 0; 
\\ \\
\text{sech}^2\left\{ - \frac{1}{2} \ln(1-m/m_*) + \frac{1}{2r_H\omega(1-m/m_*)} I_{j \ell m}\right\}  
& \text{for } 0< m < m_*;
\\ \\ 
\text{sech}^2\left\{-\frac{1}{2}\ln(m/m_*-1) + \frac{1}{2r_H\omega(m/m_*-1)}I_{j\ell m} + J^{\text{low}}_m \right\} 
& \text{for } m_* \leq m < 2m_*;
 \\ \\ 
 \text{sech}^2\left\{\frac{1}{2}\ln(m/m_*-1) + \frac{1}{2r_H\omega}I_{j\ell m} + J^{\text{high}}_m \right\}  
& \text{for } m \geq 2m_*.
\end{array}\right.
\end{equation}
Here $m_*$ is the ``critical'' azimuthal angular momentum defined by $m_* = \omega/\Omega_H$, while the quantity $I_{j\ell m}$ is defined in equation (\ref{def_Im}).

\section{Four-dimensional case $n=0$.}

When $n=0$ the Myers--Perry spacetime reduces to the usual Kerr spacetime. Furthermore,  the separation constant and effective potential reduce to those discussed in reference~\onlinecite{Kerr-Newman}. Ultimately the bounds on the greybody factors reduce (as they should) to those of reference~\onlinecite{Kerr-Newman}. 

\section{Five-dimensional case $n=1$.}

Let us now take a look at a special case with only \emph{one} extra dimension $n=1$.  These are the (3+1+1)-dimensional [five-dimensional] Myers--Perry black holes.  In this case we have the simplification
\begin{equation}
\Delta \to r^2 + a^2 -\mu.
\end{equation}
A brief computation, starting from equation (\ref{def_Im}),  now yields
\begin{equation}
I_{j\ell m}^{n=1} = \left( \frac{3}{8 a \Omega_H } -\frac{1}{8} + j^2 - \frac{a\Omega_H}{4}\right) + \left(\frac{3}{2} - j^2 - \frac{3}{8a\Omega_H} + \lambda_{j\ell m} \right)\frac{r_H}{a} \arctan \left(\frac{a}{r_H} \right).
\label{Im:n=1}
\end{equation}
Interestingly, $J_m^{\text{low}}$ has a very simple bound in five-dimensional space-time.  For $n =1$, we have
\begin{equation}
J_{m}^{\text{low}} \Big|_{n=1} \leq m\Omega_H(r_0-r_H) = \omega\frac{m}{m_*}(r_0-r_H).
\end{equation}
Let us now consider $J^{\text{high}}_{m}$; this also takes a simpler form in five-dimensional space-time
\begin{equation}
J_m^{\text{high}}\big|_{n=1} = \int_{r_0}^{\infty}\frac{m a}{2\omega}\left[ \frac{2 \omega - m \varpi(r)}{(r-r_H)(r+r_H)}\right] \,\d r.
\end{equation}
For the relevant domain of integration, $2\omega > m \varpi(r) $, then we can conclude that
\begin{equation}
J_m^{\text{high}}\big|_{n=1} \leq m a \int_{r_0}^{\infty}\frac{1}{(r-r_H)(r+r_H)}\,\d r 
= \frac{m a}{r_H} \ln\sqrt{\frac{r_0+r_H}{r_0-r_H}}.
\end{equation}
Collecting results, we finally deduce a quite explicit bound for scalar emission from five-dimensional simply rotating Myers--Perry black holes. The bound is  given by:
\begin{equation}
T_{j \ell m}^{(n=1)} \geq \left\{\begin{array}{ll}
\text{sech}^2\left\{ \frac{1}{2} \ln(1-m/m_*) + \frac{1}{2 r_H \omega} I_{j \ell m}^{n=1}\right\} 
& \text{for } m < 0;
\\ \\ 
\text{sech}^{2}\left\{\frac{1}{2r_H \omega} I_{j\ell m}^{n=1} \right\} 
& \text{for } m = 0;
\\ \\ 
\text{sech}^2\left\{ - \frac{1}{2} \ln(1-m/m_*) + \frac{1}{2r_H\omega(1-m/m_*)} I_{j \ell m}^{n=1}\right\}  
& \text{for } 0< m < m_*; 
\\ \\ 
\text{sech}^2\left\{-\frac{1}{2}\ln(m/m_*-1) + \frac{1}{2r_H\omega(m/m_*-1)}I_{j\ell m}^{n=1} + \omega \frac{m}{m_*} (r_0-r_H) \right\} 
& \text{for } m_* \leq m < 2m_*;
\\ \\ 
\text{sech}^2\left\{\frac{1}{2}\ln(m/m_*-1) + \frac{1}{2r_H\omega}I_{j\ell m}^{n=1} +  \frac{m a}{r_H} \ln\sqrt{\frac{r_0+r_H}{r_0-r_H}} \right\}  
& \text{for } m \geq 2m_*.
\end{array}\right.
\end{equation}
Here $I_{j\ell m}^{n=1}$  is as given in equation (\ref{Im:n=1}).

\section{Discussion.}

In this article we have established certain rigorous bounds on the greybody factors (mode dependent transmission probabilities) for the Myers--Perry black holes.
We have also obtained (mutatis mutandis) certain rigorous bounds on the emission rates for the super-radiant modes. In the absence of exact results,  (the relevant differential equations seem highly resistant to explicit analytic solution), quantitative bounds along these lines seem to be the best one can do. 

\section*{Acknowledgments.}

This research has been supported by a grant for the professional development of new academic staff from the Ratchadapisek Somphot Fund at Chulalongkorn University, by the Thailand Toray Science Foundation (TTSF), by the Thailand Research Fund (TRF), by the Office of the Higher Education Commission (OHEC), Chulalongkorn University, and by the Research Strategic plan program (A1B1), Faculty of Science, Chulalongkorn University (MRG5680171). 
AC was supported by the Thailand Toray Science Foundation (TTSF) and Thailand Excellence in Physics project (THEP). 
PB was additionally supported by a scholarship from the Royal Government of Thailand. 
TN was also additionally supported by a scholarship from the Development and Promotion of Science and Technology talent project (DPST). 
MV was supported by the Marsden Fund, and by a James Cook fellowship, both administered by the Royal Society of New Zealand.


\end{document}